# Recovery Time of Matter Airy Beams using the Path Integral Quantum Trajectory Model

A. L. Harris, T. A. Saxton, and Z. G. Temple

Physics Department, Illinois State University, Normal, IL, USA 61790

## Abstract

Following their discovery in the late 1970s, optical Airy beams have found numerous applications in technologies such as microscopy and optical trapping, many of which are based on the wave packets' unique features such as zero or minimal diffraction, self-acceleration, and self-healing. Recent advancements have shown that Airy beams can also be produced using matter waves with many of the same unique characteristics of their optical counterparts. We present here a study of the recovery time of damaged matter Airy wave packets in free space and a nonlinear Kerr-type medium. We show that in free space the recovery time increases approximately linearly with mass and is independent of other kinematical parameters such as momentum, velocity, and spatial width. In the Kerr-type medium, recovery time is decreased compared to free space and does not scale linearly with mass. In order to study matter Airy beams, we introduce the Path Integral Quantum Trajectory model as a new computational tool for the study of non-relativistic, quantum mechanical wave packets and demonstrate its effectiveness in dealing with heavy particle dynamics.

## 1. Introduction

There has been significant recent interest in the study and application of non-diffracting wave packets and twisted light and matter waves. With applications in fields such as micromanipulation, microscopy, and optical trapping, a fundamental understanding of the dynamics of these wave packets is crucial [1-15]. One such type of non-diffracting wave packet is the Airy wave [16]. Unlike the more traditional Gaussian wave packets, Airy waves have many unique properties that offer features not seen with their Gaussian counterparts.

The first studies and applications of Airy wave packets have been in optical contexts [17-28], where they have been shown to be non-dispersive and exhibit force-free self-acceleration and self-healing properties. At first glance, some of these properties might seem too good to be true, however careful analysis shows that these features are real. For example, Airy beams are non-dispersive in the same sense that plane waves are non-dispersive. In their exact form, they are not square normalizable and extend throughout all space. They carry an infinite amount of energy [24] and its intensity remains constant throughout time. In reality, any Airy beam produced in the laboratory will be localized and therefore is square normalizable. This localization does cause the wave packet to diffract, although individual lobes within the beam do so at a slower rate than their Gaussian counterparts. The force-free acceleration seems to defy Ehrenfest's theorem, but while it is true that the individual spatial peak structures within the wave packet follow parabolic trajectories, the expectation value of a truncated Airy beam does not exhibit acceleration.

In addition to their propagation through free space, the propagation of optical Airy beams through nonlinear media has also been examined. In these media, Airy beams exhibit qualitatively different behavior, including features such as self-focusing [29-32]. Their stability over time is altered compared to similar wave packets in free space [33], and two interacting Airy beams have been shown produce spatial solitons [32].

While the vast majority of studies involving Airy wave packets have been in the optical context, recent experimental developments have demonstrated that electron Airy beams can now be produced in the lab [34]. These newly created matter Airy waves have many of the same properties as their optical counterparts and have the potential for many new and interesting applications, such as control and rotation of nanoparticles [6-11,19,35], improved resolution in electron microscopy [9,10,21,22], and the steering of electronic wave packets [34]. However, due to their recent discovery, matter Airy waves are much less understood.

Here we focus on the self-healing feature of Airy beams and define a recovery time for a damaged beam to recover its structure. The self-healing of an Airy beam occurs when a portion of the beam is perturbed or encounters an obstacle. Continued propagation of the beam transfers energy from the tail to the head of the beam, resulting in the main peak being restored, although at a decreased magnitude. We examine the propagation of damaged Airy waves in both free space and a nonlinear Kerr-type medium. We show that in both cases the recovery time increases as particle mass increases, and for wave packets in free space, it is independent of kinematical parameters. We also show that damaged Airy waves recover faster in a nonlinear medium than in free space.

With the recent creation of electron Airy beams, and the possibility for more massive Airy beams around the corner, it is essential to understand their unique properties at a fundamental level if the numerous proposed applications are to be achieved. To our knowledge, this work represents one of the first analyses of the self-healing properties of matter Airy beams in free space and in a nonlinear medium, and we hope that it spurs further theoretical and experimental developments in this area.

In order to study the dynamics of matter Airy beams, we introduce the Path Integral Quantum Trajectory (PIQTr) model as a new computational technique for finding the time evolution of particle wave functions and demonstrate the model's effectiveness in treating heavy particles. Because the PIQTr model does not rely on any classical, semi-classical, saddle point, or other approximation, it provides a highly accurate computational tool for the calculation of time-dependent wave functions and is especially useful for heavy particle dynamics.

Traditional approaches for studying the quantum mechanical motion and interactions of non-relativistic charged particles often focus on solving the Schrödinger equation. In the time-dependent case, a variety of numerical methods are available, including finite difference methods, spectral representations, and finite element models. However, one of the challenges with both time-independent and time-dependent quantum mechanical methods is the computational difficulties encountered as the mass of the particle increases. In particular, methods based on basis expansions often require a large number of basis states, making the calculation difficult, if not impossible. Alternatively, matrix techniques can require the manipulation of large matrices, which is computationally challenging. In order to overcome these challenges, many methods employ classical or semi-classical treatments to make the calculation manageable.

One technique that is often overlooked when it comes to solving non-relativistic quantum mechanical dynamics is the Feynman path integral method, which has been widely used in high energy physics, quantum field theory, and statistical mechanics. However, it is not always a researcher's first instinct for fundamental physics problems [36-44]. Based on a Lagrangian approach, the formal path integral technique yields results completely equivalent to the Hamiltonian-based Schrödinger equation approach [45]. However, one of the practical difficulties associated with the method is the infinite sum over paths (or amplitudes associated with paths) required in the calculation. Our PIQTr model overcomes the need for this infinite sum by taking advantage of a heavy particle's small deBroglie wavelength in order to greatly reduce the computational requirements. The theoretical and numerical details of the PIQTr model are included in section 2, with a detailed analysis of the computational requirements presented in the appendices. Section 3 is dedicated to the discussion of matter Airy beam recovery time and in Section 4 a general summary and conclusion are given. Atomic units are used throughout unless otherwise stated.

## 2. PIQTr Model

The PIQTr model introduced here can be used to calculate time-dependent wave functions for a particle moving in any one-dimensional potential without approximation. This potential may or may not be time dependent. Fundamental quantum mechanics dictates that the time evolved wave function can be found by applying the propagator (also called the kernel or time evolution operator) to the initial state wave function

$$\psi(x_b, t_b) = \int_{-\infty}^{\infty} K(x_a, x_b, t_a, t_b)\psi(x_a, t_a)dx_a, \qquad (1)$$

where $\psi(x_a, t_a)$ is the initial state wave function at time $t_a$, $K(x_a, x_b, t_a, t_b)$ is the propagator and $\psi(x_b, t_b)$ is the new time evolved wave function at time $t_b$.

The difficulty in the practical application of Eq. (1) is knowing the propagator. In the path integral method, the propagator is written as a sum over amplitudes for all trajectories for the particle to travel from its initial position to a final position. Each trajectory of the particle contributes an amplitude equal in magnitude, but different in phase

$$\phi[x(t)] = \frac{1}{A}\exp\left(\frac{iS[x_b, x_a]}{\hbar}\right), \qquad (2)$$

where $S[x_b, x_a]$ is the classical action for the trajectory $x(t)$ from $x_a$ to $x_b$. It is given by

$$S[x_b, x_a] = \int_{t_a}^{t_b} L(x, \dot{x})dt, \qquad (3)$$

where $L$ is the Lagrangian of the system

$$L(x, \dot{x}) = \frac{1}{2}m\dot{x}^2 - V(x, t) \qquad (4)$$

for a particle of mass *m* moving in a potential *V(x,t)*. The constant *A* is the normalization constant. Combining Eqs. (1) and (2), the propagator can be expressed as

$$K(x_b, x_a, t_a, t_b) = \int \phi[x(t)] \mathcal{D}x(t), \tag{5}$$

where $\mathcal{D}x(t)$ represents an integral over all trajectories. Inserting Eq. (5) into Eq. (1) yields

$$\psi(x_b, t_b) = \int \int \phi[x(t)] \psi(x_a, t_a) \mathcal{D}x(t) dx_a. \tag{6}$$

At this point, the details of the PIQTr model become clearer by using a discrete formalism, with the added advantage of a direct connection to the computational method.

We begin by discretizing space and time, assuming for simplicity in this derivation uniform grids for both. Then,

$$\psi(x_b, t_b) = \sum_{x_a} \sum_{\substack{paths \\ x(t)}} \phi[x(t)] \eth x(t) \psi(x_a, t_a) \Delta x_a, \tag{7}$$

where ðx(t) represents an infinitesimal difference in paths.

Using Eq. (2), this becomes

$$\psi(x_b, t_b) = \sum_{x_a} \sum_{\substack{paths \\ x(t)}} \frac{1}{A} e^{\frac{iS[x(t)]}{\hbar}} \eth x(t) \psi(x_a, t_a) \Delta x_a. \tag{8}$$

Consider now the evolution of the wave function through only one time step, such that $\psi(x_b, t_b) = \psi(x_b, t_a + \Delta t)$. Then, assuming that the time step is infinitesimally small, there is only one path between two spatial points, which is a straight line. In this case, the propagator becomes

$$K_{1step}(x_b, x_a, t_a, t_a + \Delta t) = \frac{1}{A_{1step}} e^{\frac{iS_{1step}}{\hbar}}, \tag{9}$$

with $S_{1step}$ given by

$$S_{1step} = L_{1step} \Delta t, \tag{10}$$

where $L_{1step}$ is the Lagrangian at time $t_a$ and position $x_a$. In one dimension, the Lagrangian for a single time step is given by

$$L_{1step} = \frac{m}{2}\frac{(x_b-x_a)^2}{\Delta t^2} - V(x_a, t_a). \tag{11}$$

The normalization constant in Eq. (9) is given by $A_{1step} = \sqrt{\frac{2\pi i\hbar\Delta t}{m}}$ [1].

Combining Eqs. (8) – (11) yields the evolution of the initial state wave function by one infinitesimal time step

$$\psi(x_b, t_a + \Delta t) = \sum_{x_a} \frac{1}{A_{1step}} e^{\frac{iS_{1step}}{\hbar}} \psi(x_a, t_a) \Delta x_a. \tag{12}$$

In integral form, this is

$$\psi(x_b, t_a + \Delta t) = \int_{-\infty}^{\infty} \frac{1}{A_{1step}} e^{\frac{iS_{1step}}{\hbar}} \psi(x_a, t_a) \mathrm{d}x_a. \tag{13}$$

Equation (12) or (13) now allows for the iteration of the wave function by replacing $\Psi(x_a, t_a)$ with $\Psi(x_b, t_a + \Delta t)$. Successive applications will result in the time evolved wave function at any desired final time.

The derivation above is exact assuming that the time step $\Delta t$ is small enough to approximate the limit that $\Delta t \to 0$. This requirement comes from the use of Eq. (10) and the assumption that for infinitesimal $\Delta t$ there is only one straight-line path between two neighboring spatial points. A detailed discussion of the practical requirements for $\Delta t$ can be found in Appendix A.

Brute force numerical evaluation of the integral in Eq. (13) is possible, but cumbersome due to the presence of $K_{1step}$, which for fixed $x_b$ and $\Delta t$ is an oscillatory function of $x_a$. These oscillations increase as the magnitude of $S_{1step}$ increases, which occurs for small $\Delta t$, large mass, or large differences between $x_a$ and $x_b$. However, it is possible to take advantage of the oscillatory nature of $K_{1step}$ when performing the integral. The principle of least action dictates that the classical path of the particle is the one that minimizes the action. Therefore, paths that deviate significantly from the classical path will have large values of the action, resulting in the single step propagator being a highly oscillatory function of $x_a$. Assuming that the wave function is smooth over the wavelength of these rapid oscillations in the propagator, the integrand of Eq. (13) is as likely to be positive as it is negative, and contributions to the sum of paths far from the classical path will effectively be zero [45]. This implies that the integration range for Eq. (13) does not need to encompass all of space, but rather a small region in which the single step propagator has minimum oscillations. The required integration range $\delta$ can be shown analytically to scale as $\delta \sim \sqrt{\frac{\Delta t}{m}}$, and numerical results agree with this to within a factor of 2 (see Appendix B for details).

The theoretical development above assumes constant temporal and spatial grids, and in practice, we use a constant time grid with $\Delta t = \frac{t_b - t_a}{N_t - 1}$ where $N_t$ is the number of time grid points. However, the two spatial grids for the integration and the iterated wave function are not uniform and result from the Gaussian quadrature points used for the integration.

The runtime of the code scales linearly with $N_t$ and as the product of the number of spatial integration points $N_{xint}$ and the number of wave function grid points $N_{xwf}$. In general, the calculation times are quite manageable on a single processor, however parallelization of the integration is straightforward and serves to further improve runtime, scaling linearly with number of processors. A detailed analysis of the computational requirements of the PIQTr model can be found in Appendix C.

### 3. Results

As noted above, Airy beams were first found to be solutions to the free particle Schrödinger equation in 1979 [16], but only in the last decade have they received significant attention. Much of the focus has been within optical contexts [17-28] following the experimental and theoretical investigations of [17,28]. More recently, the production of electron Airy beams was demonstrated by passing electrons through a holographic mask that imparts a cubic phase [34]. These matter Airy beams exhibited the same unique properties of optical Airy beams, including curved trajectories, self-healing, and limited diffraction.

*3.1 Free Particle Propagation*

We begin with an analysis of free particle Airy wave packets. The discovery of Airy beams as non-diffracting particles was due to the relationship between the time-dependent free particle Schrödinger equation

$$\frac{i}{\hbar}\frac{\partial \psi}{\partial t} = -\frac{\hbar^2}{2m}\frac{\partial^2 \psi}{\partial x^2} \tag{14}$$

and the paraxial wave equation

$$i\frac{\partial \psi}{\partial z} = -\frac{1}{2}\frac{\partial^2 \psi}{\partial x^2}, \tag{15}$$

whose solution was known to be Airy functions [16]. The exact solution to Eq. (14) can be written as [16]

$$\psi(x,t) = Ai\left[\frac{B}{\hbar^{\frac{2}{3}}}\left(x - \frac{B^3 t^2}{4m^2}\right)\right] e^{i\left(\frac{B^3 t}{2m\hbar}\right)\left(x - \frac{B^3 t^2}{6m^2}\right)}, \tag{16}$$

where $B$ is an arbitrary parameter (taken to be 1 here). From the $t^2$ term in the Airy function, it is easy to see that this wave will exhibit an acceleration of $\frac{1}{2m^2}$ [16]. In our investigations, we use an initial state, exponentially truncated Airy beam with the leading peak centered at $x = 0$ and general form [18]

$$\psi(x,0) = Ai[x]e^{ikx}e^{\gamma x}. \tag{17}$$

The truncation parameter $\gamma$ determines the width of the initial Airy wave packet. A larger value of $\gamma$ results in a more narrow initial wave packet. Because the acceleration of the Airy beam peaks is known, the classically predicted trajectory is given by

$$x(t) = x_0 + v_0 t + \frac{1}{2}at^2 \tag{18}$$

with $a = \frac{1}{2m^2}$. Clearly, as the mass of the particle increases, the acceleration is reduced.

Figure 1 shows the time evolution of free particle Airy beams with truncation parameter $\gamma = 0.1$, initial velocity $v_0 = 1$ and masses $m = 1, 10,$ and $100$ a.u. With time on the vertical axis and space on the horizontal axis, the color scale represents the probability density $|\psi(x,t)|^2$. The decreasing acceleration with increasing mass is apparent with the trajectories for heavier particles having less curvature. It can also be seen that the Airy wave packet maintains its shape and magnitude for a finite time. Over time, the Airy wave packet spreads, leading to a decrease in the truncation parameter for $t > 0$. Calculations show that the truncation parameter is halved at a time that is directly proportional to mass. Therefore, the stability of an Airy beam's shape over time is improved for more massive particles.

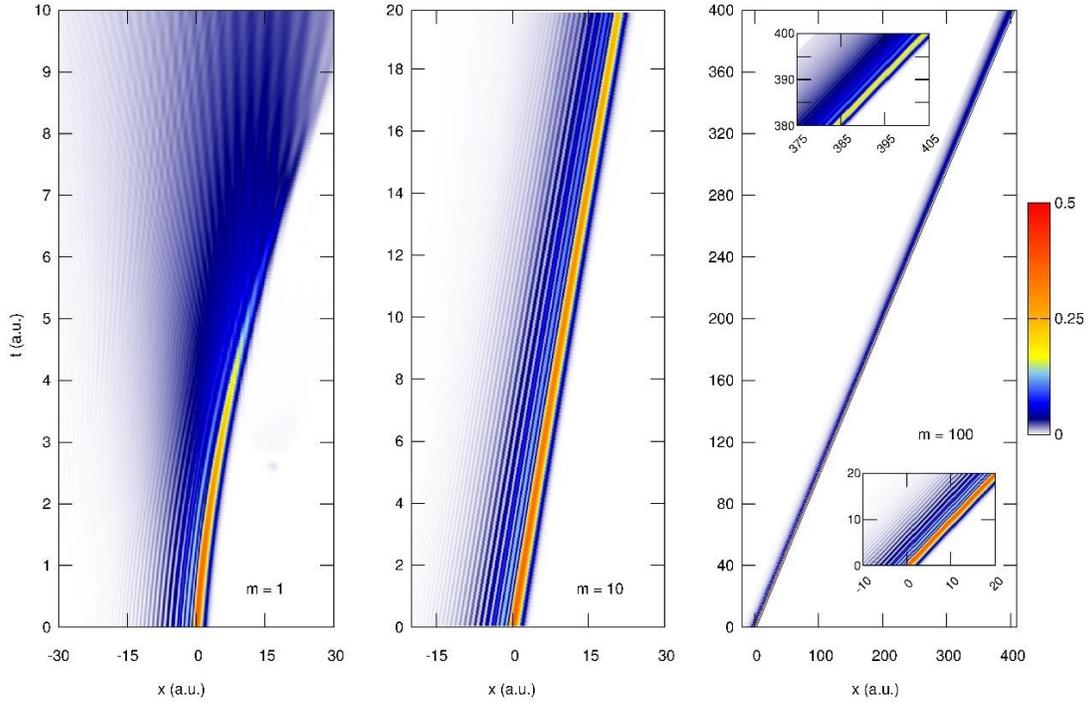

**Figure 1** Time evolution of free particle Airy beams with masses 1, 10, and 100 a.u. and truncation parameter $\gamma = 0.1$. The color scale is the magnitude of the probability density $|\psi(x,t)|^2$. The insets in the third panel show a more detailed view of the short- and long-time regimes.

The primary feature of the matter Airy beams in which we are interested is their self-healing nature. When the Airy beam is perturbed or damaged due to an obstacle, the beam will recover its initial structure after some time. This has been demonstrated extensively with optical Airy beams [17,18,23,24], and shown to hold true with electron Airy beams [34]. Given the recent intense interest in optical Airy beams and the creation of electron Airy beams, it is worthwhile to determine how the self-healing properties change with particle mass. In particular, because any practical application of matter Airy beams will involve their interaction with matter, it is important to quantify the time it takes for the wave packet to recover its shape following some distortion or damage to the original wave packet. For example, if an Airy wave packet is perturbed through a scattering event or truncated by a collimating slit, it will be useful to know if the wave packet will recover its shape before a second interaction might occur. In particular, we study Airy beams whose initial leading peak (in the direction of travel) has been blocked by an obstacle, as might be encountered with a collimating slit for a scattering experiment. For comparison, Fig. 2 shows initial state damaged and undamaged Airy beams. Both are normalized such that $\int |\psi(x,0)|^2 = 1$.

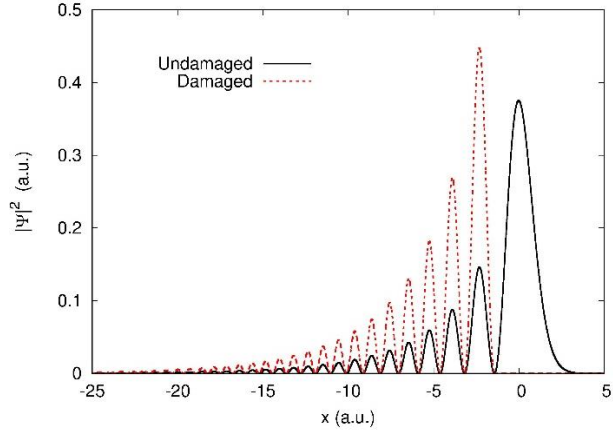

**Figure 2 Initial state damaged and undamaged Airy beams for m = 1, $\gamma = 0.1$, and p$_0$ = 1. Both wave functions are normalized to unity.**

In order to quantify the time it takes an Airy beam to self-heal, we introduce a definition of '*recovery time*', which we define as the time it takes for the location of the leading peak maximum in the damaged Airy beam to reach the predicted location of the leading peak maximum in an undamaged Airy beam. In other words, the time for the trajectory of the damaged Airy beam's leading peak to reach the predicted path of an undamaged Airy beam's leading peak is defined as the recovery time.

This definition of recovery time presents a practical challenge because the damaged Airy beam's leading peak approaches the predicted location gradually in an asymptotic-like fashion. Thus, a specific time for recovery is difficult to define. However, a general trend of how recovery time scales with mass can be found. We determine the recovery time by taking the ratio of the predicted leading peak location to the leading damaged peak's location. When this ratio is within 1% of unity, we consider the damaged Airy beam to be fully recovered. Fig. 3 shows a sample plot of the ratio of leading peak locations, as well as the leading peak trajectory of the damaged Airy beam and the corresponding undamaged trajectory (m = 1, $p_0$ = 1, $\gamma$ = 0.1). The discontinuity in the ratio occurs at the time where the damaged Airy beam's leading peak location changes from a negative to a positive value. The undamaged leading peak's position is always positive. At long times, the peak location ratio approaches unity as the damaged Airy beam returns to its original trajectory and remains one for further propagation. This indicates that the Airy beam has recovered and will follow its original trajectory indefinitely. The blue arrows shown in Fig. 3 point to the recovery time when the peak location ratio is within 1% of unity.

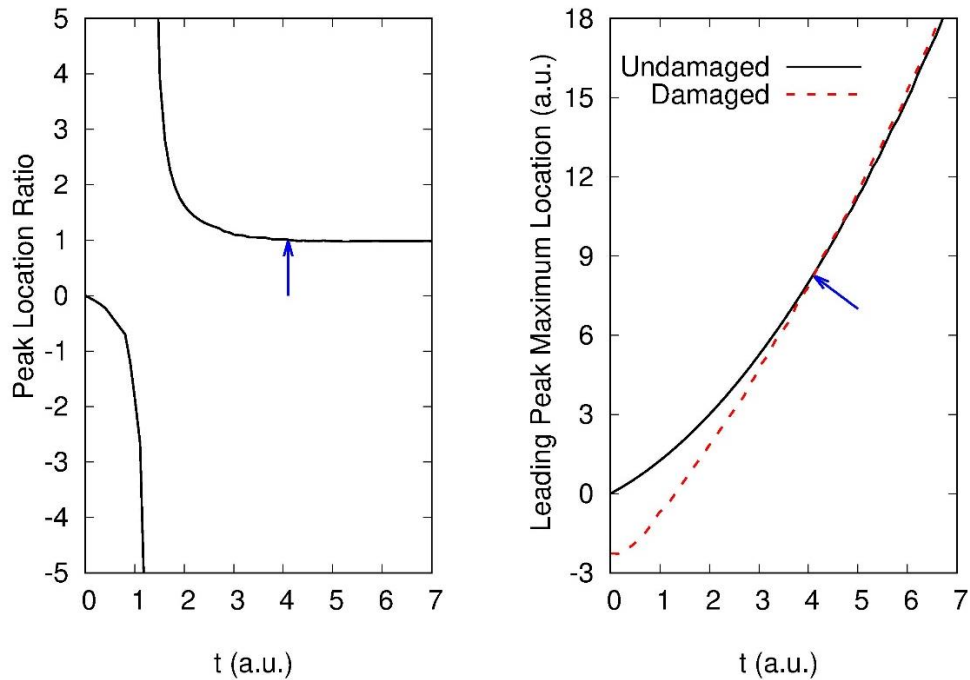

**Figure 3 Left panel – ratio of undamaged wave function leading peak maximum location to damaged wave function leading peak maximum location. Right panel – trajectories for leading peak maxima for damaged and undamaged Airy wave packets. Results are shown for m = 1, $p_0$ = 1, and $\gamma = 0.1$. Blue arrows point to recovery time ($\tau_r = 4.1\ a.u.$).**

Figure 4 shows the free particle time evolution of three damaged Airy beams with truncation parameter $\gamma = 0.1$. Each of the Airy beams has been damaged by removing the leading peak (see Fig 2). Results are shown for particles with $(m, p_0) = (1,1)$, $(10,10)$, and $(1,10)$. Figs 4 (a) and (b) show results for particles with the same velocity, but different masses. A comparison of these two cases shows that the recovery time varies with mass, despite the same particle velocity. Figs 4 (a) and (c) show results for particles with the same mass, but different velocities and momenta, and it can be seen that the recovery time does not change as the initial velocity changes. In addition to varying the mass, momentum, and velocity, we also performed calculations for Airy beams with a larger truncation parameter and found that the recovery time was approximately independent of width, as shown in Table 1.

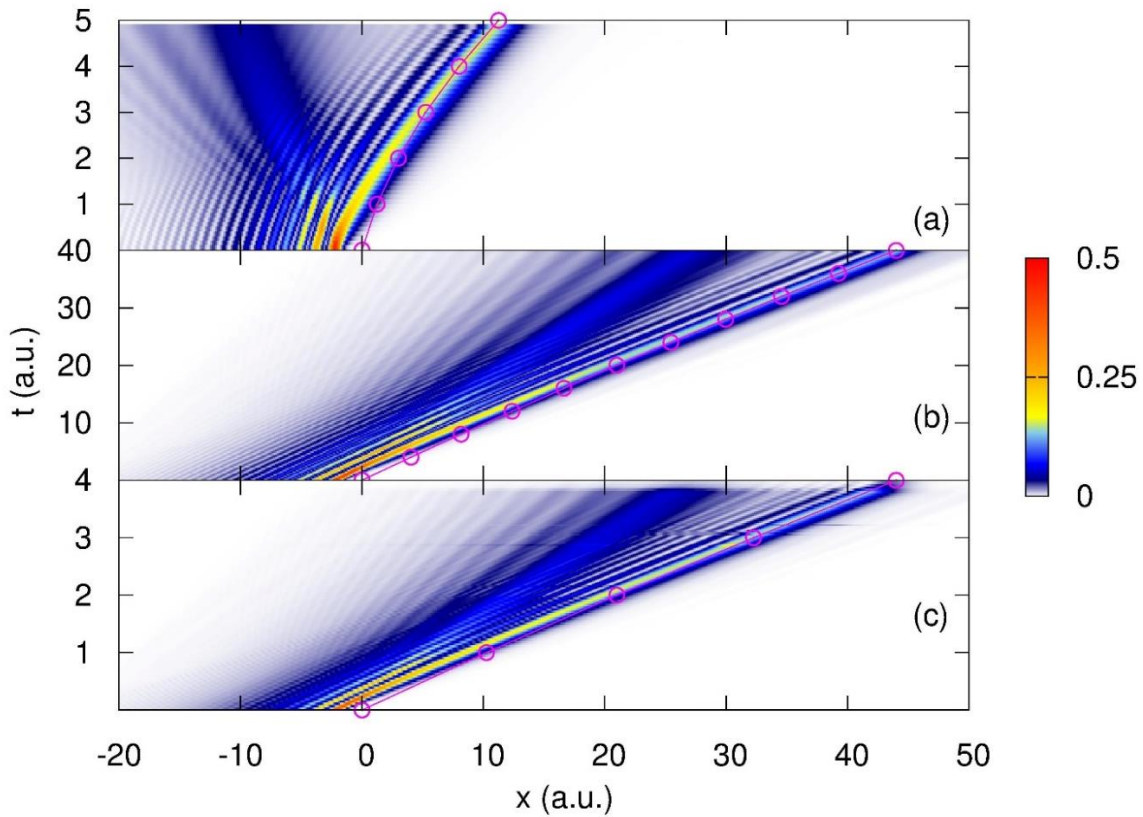

**Figure 4** Time evolution of damaged free particle Airy beams with (a) m = 1, $p_0$ = 1, (b) m = 10, $p_0$ = 10, (c) m = 1, $p_0$ = 10. The initial Airy beam's leading peak was removed and the wave packet was allowed to evolve freely. The color scale is intensity of the probability density $|\psi(x,t)|^2$. The pink line with circles is the trajectory of the leading peak of an undamaged Airy beam.

As can be seen from Figs 3 and 4, the process of self-healing evolves such that the leading peak in the damaged wave function gradually returns to the trajectory of the leading peak for the undamaged wave function (shown as the pink line with circles). Quantitative results for recovery times are shown in Table 1, and it can be seen that for free particle propagation the recovery time scales approximately linearly as four times the mass of the particle. In general, all of our testing indicates that the recovery time is primarily dependent upon the mass of the particle and is independent of kinematical parameters such as momentum, velocity, and truncation. The approximately linear scaling of the recovery time with mass and the insensitivity to momentum, velocity, and initial width implies that more massive Airy beams, such as those created with muons or protons, will take a very long time to recover. This could have important implications on any applications that might rely on the self-healing properties of Airy beams. For heavy particle damaged waves, recovery will likely not occur during the time scales of an experiment.

|        | $\gamma = 0.1$   | $\gamma = 0.4$ |
| ------ | ---------------- | -------------- |
| m = 1  | $\tau_r = 4.1$   | 4.1            |
| m = 5  | 20.8             | 20.1           |
| m = 10 | 39.3             | 35.6           |
| m = 20 | 72.6             | 77.6           |

**Table 1 Recovery time of damaged Airy beams with initial velocity of 1 a.u., but different masses and truncation parameters.**

*3.2 Nonlinear Propagation*

The studies of optical Airy beams in free space have been expanded to their propagation and interaction in nonlinear media, such as a Kerr medium in which the presence of an electric field alters the medium's index of refraction [32,33,46-48]. Propagation of a single optical Airy beam in a focusing Kerr medium results in an enhancement of the second lobe intensity [33], but the leading lobe follows the same trajectory as in free space. The nonlinearity can also affect the stability of the Airy beam [48-50] and interacting beams have been shown to generate spatial solitons [32]. Additional features such as self-focusing or trapping of optical Airy beams in nonlinear media [51] have also been observed.

In general, matter Airy beams have been studied much less than those in optical contexts, and the same is true for their propagation in nonlinear media. While the Kerr effect is an optical response, there is evidence for the propagation of matter waves in nonlinear media. For example, the nonlinear Schrödinger equation with a cubic interaction term (as in the Kerr effect) can be used to describe some Bose Einstein condensates [52,53].

Here we investigate matter Airy beams moving in a nonlinear focusing Kerr-type medium where the wave function evolves according to the nonlinear Schrödinger equation

$$\frac{i}{\hbar}\frac{\partial \psi}{\partial t} = -\frac{\hbar^2}{2m}\frac{\partial^2 \psi}{\partial x^2} + \delta n \psi, \qquad (19)$$

where $\delta n$ is given by $g|\psi(x)|^2$ and $g$ is a nonlinearity constant. In an optics context, $\delta n$ represents a variable index of refraction, while for matter Airy beams, $\delta n$ is a mean field term that is proportional to particle density [53].

As in the optics case, the propagation of matter Airy beams in the Kerr-type medium is qualitatively different than that of free particles. Figure 5 shows a comparison of undamaged free space Airy waves to those in the Kerry-type medium for m =1 and m = 10. The matter Airy wave in the nonlinear medium shows an enhancement in the secondary peaks at small times for both masses. As the beams are allowed to evolve, the leading peak in the nonlinear medium spreads and loses intensity over time, unlike the wave packet in free space. Also, in the Kerr-type medium, the leading peak maximum follows a similar trajectory to that of an Airy wave packet in free space, but lags slightly behind (see Figs 5 and 6). This decreased acceleration could be due to the self-focusing that occurs in nonlinear media [48], where energy intensifies each lobe, in addition to accelerating it. Also, some interference structures can be observed in the tail of the wave function at longer times. These structures result from negative momentum components that enter into the wave function as a result of the nonlinear medium.

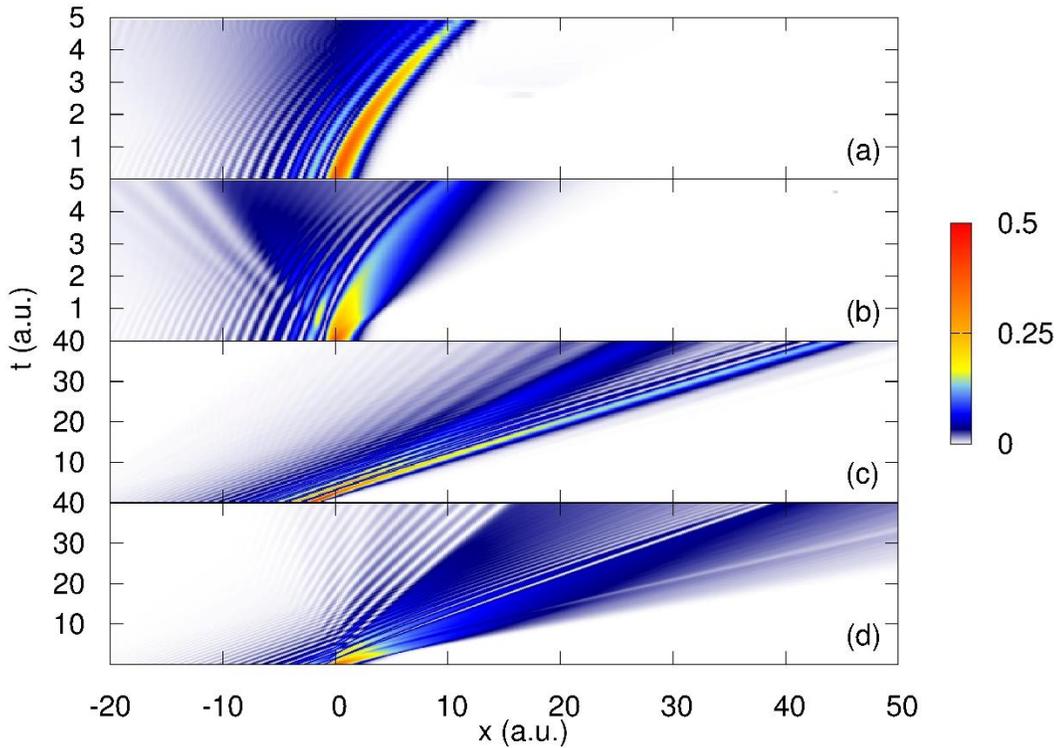

**Figure 5** Time evolution of undamaged Airy beams with $\gamma = 0.1$ and (a,b) m = 1, $p_0$ = 1 and (c,d) m = 10, $p_0$ = 10. In (a,c) the Airy beams evolve in free space, while in (b,d) they evolve in the nonlinear Kerr-type medium with nonlinearity constant g = 10. The color scale is intensity of the probability density $|\psi(x,t)|^2$.

Next, we turn to the calculation of recovery time for the Airy beam in a Kerr-like medium. We use the same definition of recovery time that was introduced above and note that the dynamics of the evolving damaged Airy beam in the nonlinear medium are very similar to those of the damaged Airy beam in free space. However, the evolution of the undamaged Airy beam in the nonlinear medium is different from its free space counterpart. This leads to differences in the recovery times. In particular, the damaged Airy beam no longer asymptotically approaches the undamaged beam trajectory in the Kerr-type medium. Instead, the damaged Airy beam leading peak trajectory approaches and then crosses the undamaged trajectory, asymptotically approaching the undamaged free space trajectory, as seen in Fig 6. The results in Fig 6 are shown for $m = 1$ and $p_0 = 1$, but results for other mass particles show similar behavior.

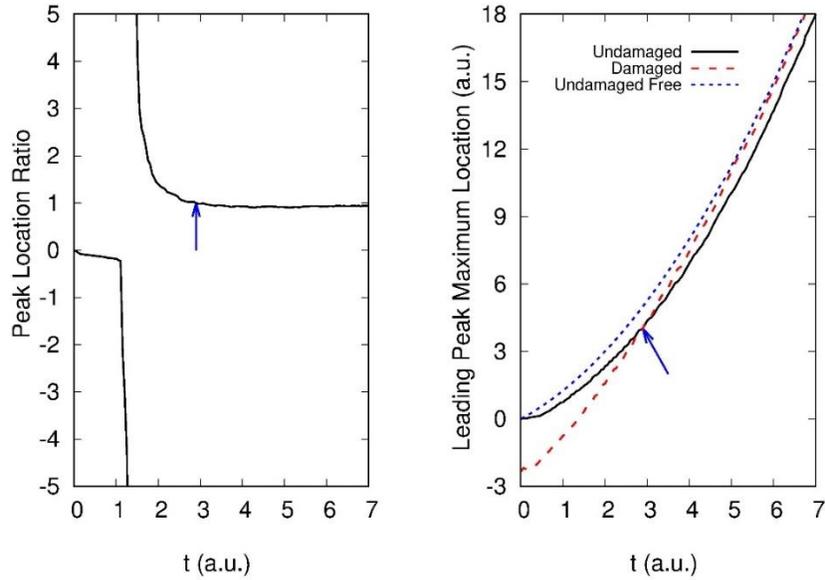

**Figure 6 Same as Fig. 3, but for Airy wave packets moving in nonlinear Kerr-type medium with $g = 10$. Results are shown for m = 1, p₀ = 1, and $\gamma = 0.1$. Blue arrows point to recovery time ($\tau_r = 2.9\ a.u.$).**

Because the undamaged beam accelerates more slowly in the nonlinear medium, the damaged Airy beam recovers more quickly than it does in free space, with recovery times 25 to 50% faster (see Table 2).

|  | Free space | Nonlinear medium |
|---|---|---|
| m = 1 | $\tau_r = 4.1$ | 2.9 |
| m = 5 | 20.8 | 10.5 |
| m = 10 | 39.3 | 24.2 |
| m = 20 | 72.6 | 32.5 |

**Table 2 Recovery time of damaged Airy beams with initial velocity of 1 a.u., $\gamma = 0.1$, and different masses for propagation in the nonlinear Kerr-type medium and free space.**

# 4. Conclusion

In summary, we examined the self-healing property of damaged Airy beams in free space and a nonlinear Kerr-type medium. Damage was induced by removing the leading peak to simulate an encounter with an obstacle. A recovery time for the wave packet was defined and shown to increase approximately linearly with increasing particle mass in free space. Additionally, in free space, this recovery time was independent of both velocity, momentum, and width of the wave packet. In the nonlinear medium, the recovery time also increased with particle mass, but not linearly. A comparison between recovery time in free space and recovery time in the nonlinear medium showed that the Airy beam recovers more quickly in the nonlinear medium. This decreased recovery time could be due to the self-focusing aspects of Airy beam dynamics in nonlinear media. The increase in recovery time with mass will have important implications in any future experiments involving massive Airy beams and indicates that even in nonlinear media, the heavy Airy beams will be unlikely to self-heal during the time frame of an experiment.

The time-dependent Airy beam wave functions were calculated with our newly introduced Path Integral Quantum Trajectory (PIQTr) model and the numerical algorithm of the method was introduced. By taking advantage of the smaller deBroglie wavelength of heavy particles, the PIQTr model scales favorably with increasing mass. This feature will be useful for future applications to heavy particle dynamics and lays the foundation for the PIQTr model being one of the first that can be applied to both heavy and light particles with similar computational requirements.

## Appendix A

The theoretical derivation of the PIQTr model requires that the time step for iterations of the wave function be infinitesimally small such that there is only one linear path between two spatial points. Then, the propagator can be written as Eq. (9). The corresponding numerical requirement is that $\Delta t$ is sufficiently small that the propagated wave function will not change as Δt decreases. Experience shows that if Δt is too large, the wave function will propagate too quickly. However, if it is too small, the computation becomes intensive. The key to achieving accurate results is to keep Δt as small as possible without resulting in an unnecessary computational burden. Unfortunately, there is no way to analytically approximate Δt because it will depend on the potential chosen, as well as the mass.

We can gain, however, insight into the requirements of $\Delta t$ by studying the familiar Gaussian wave packet moving under the influence of a constant force. In this case, the analytical result is well-known, which allows for detailed testing of the effect of Δt on the time propagation of the wave function. Other systems will have similar, but not identical, requirements. The force is chosen to be 2 a.u. and the mass is chosen to be 1 a.u. The initial state wave function is a Gaussian with σ = 1, no initial momentum ($p_0 = 0$), and initial position $x_0 = 0$ a.u.

$$\psi(x,0) = \frac{e^{-(x-x_0)^2/2\sigma^2}}{(\pi\sigma^2)^{1/4}} \tag{A1}$$

Fig. A1 shows the time sequence for the propagated wave function with results for different values of Δt compared to the analytically exact answer, and Table A1 shows the expectation value for the final time ($t = 20$ a.u.) compared to the analytically exact answer.

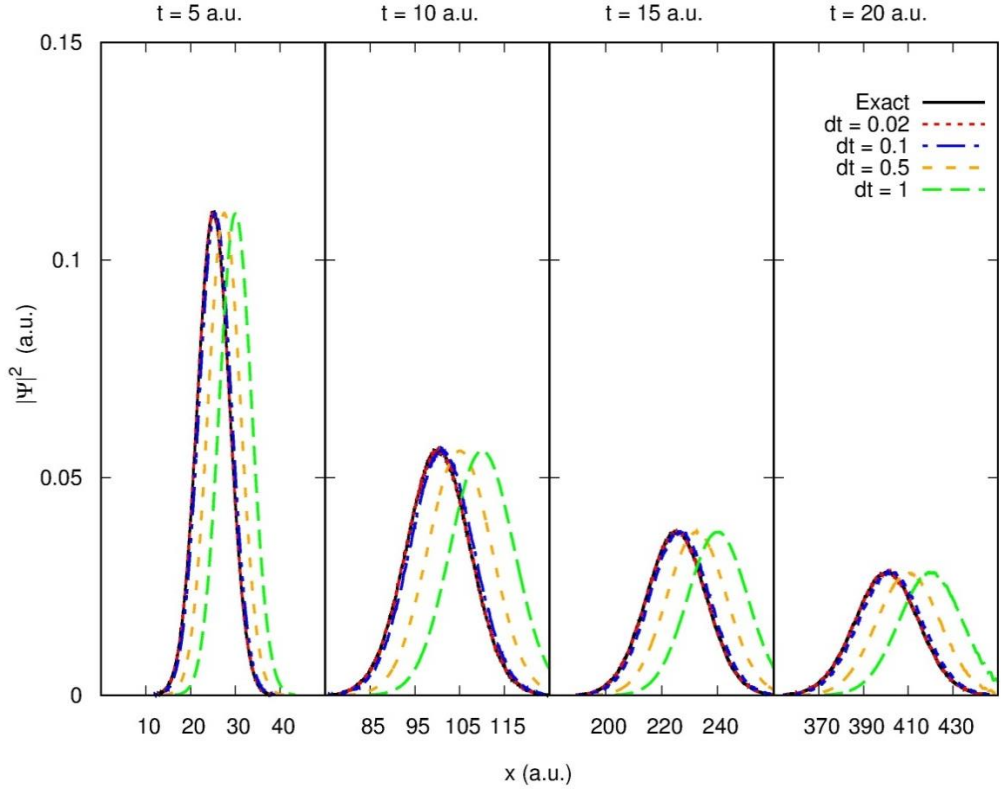

**Figure A1** Time sequence for Gaussian probability density moving under a constant force with mass = 1 a.u., force = 2 a.u., σ = 1, and zero initial momentum. Results are shown for different size time steps Δt compared to the analytically exact answer.

| Δt | <x> | % difference from exact |
|---|---|---|
| 0.02 | 400.4 | 0.2% |
| 0.1 | 402.0 | 0.5% |
| 0.5 | 408.2 | 2.1% |
| 1 | 419.9 | 5.0% |

**Table A1** Expectation value of the wave function at the final time of t = 20 a.u. for a particle of mass 1 moving under a constant force of 2 a.u. The initial position of the wave function was $x_0 = 0$, and the analytically exact expectation value at t = 20 a.u. is 400 a.u.

From Fig A1 and Table A1, we can see that when *Δt* is too large, the wave packet propagates with too large of a speed. This is due to a shift in the peaks of the single step propagator compared to the analytically exact propagator. We show in Fig. A2 a plot of the single step propagator $K_{1step}$ as a function of $x_a$ compared to the analytically exact answer $K_{exact}$ for the Δt values shown in Fig. A1. The value of $x_b$ was fixed at 0.

From Fig. A2, it is easy to see that as the time step decreases, the agreement between the exact answer and the numerical result improves. When $\Delta t$ is too large, the $K_{1step}$ peaks are shifted to larger $x_a$ from $K_{exact}$. Also, as $\Delta t$ increases, the peaks of $K_{exact}$ shift away from zero. Therefore, a $\Delta t$ value that is too large effectively gives peaks in locations similar to those of $K_{exact}$ at a later time, which results in an increase in velocity.

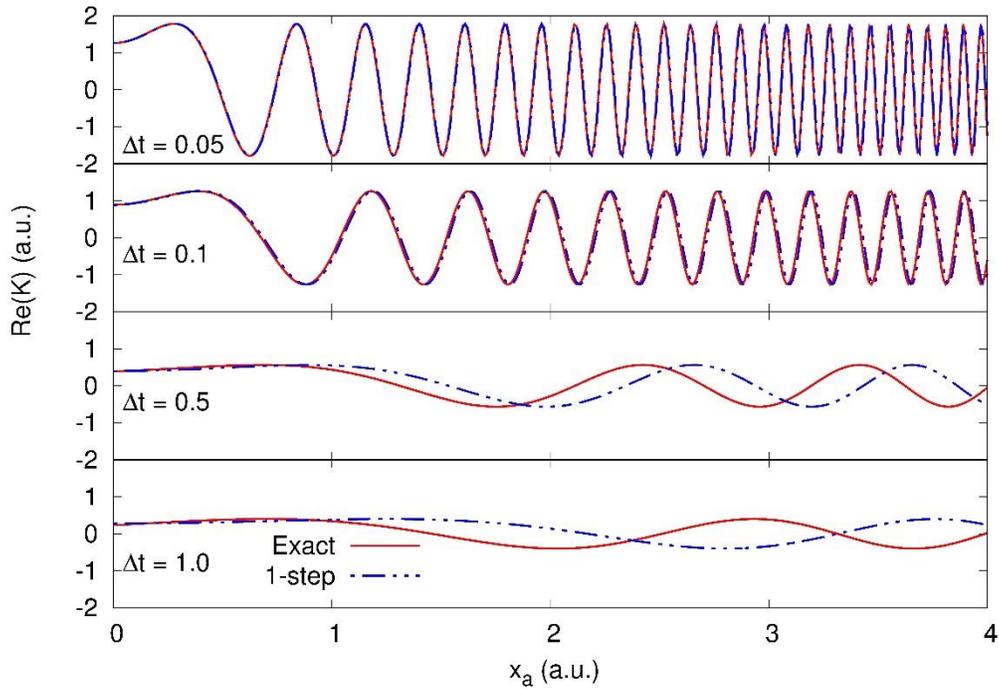

**Figure A2 Real part of the single step propagator for a particle of mass 1 a.u. moving under a constant force of F = 2 a.u. Results for different values of $\Delta t$ are compared to the analytically exact propagator. The results are plotted as a function of $x_a$ for fixed $x_b = 0$.**

Finally, in Fig. A2, one also sees that as the time step increases, fewer oscillations are observed in the propagator over the same spatial distance. This is easily predicted from Eqs. (10) and (11), where the action is seen to vary inversely with $\Delta t$ in the first term and linearly with $\Delta t$ in the second term. Thus, as $\Delta t$ increases, there are fewer oscillations over the same distance, indicating that a particle may move more slowly in order to arrive at the same $x_b$ point at a later time.

## Appendix B

In its exact form, the PIQTr model requires a numerical integration over all space, however in practice the finite deBroglie wavelength of a particle allows for this integration region $\delta$ to be reduced. To determine the required numerical integration region, note that $K_{1step}$ is centered around $x_a = x_b$. When $x_a$ is sufficiently far from $x_b$, the integrand becomes highly oscillatory and the positive and negative contributions to the integral will cancel. Therefore, for $|x_a - x_b| \geq \delta$, there is no need to perform the integration and Eq. (13) becomes

$$\psi(x_b, t_a + \Delta t) = \int_{x_b-\delta}^{x_b+\delta} K_{1step}(x_b, x_a, t_a, t_a + \Delta t)\, \psi(x_a, t_a) dx_a. \tag{B1}$$

This results in a significant computational advantage due to the much smaller integration range.

A semi-empirical estimate for $\delta$ can be found from examining how the wavelength of the single step propagator changes with increasing $|x_a - x_b|$. Consider the free particle single step propagator in which $V(x_a, t_a) = 0$ in Eq. (11). The de Broglie wavelength can be written as

$$\lambda = \frac{h}{p} = \frac{2\pi\hbar\Delta t}{m(x_b - x_a)}. \tag{B2}$$

We assume that it is safe to neglect the contributions to the integral when the change in wavelength as a function of $x_a$ is less than some value $\beta$. Therefore,

$$\frac{d\lambda}{dx_a} = \frac{2\pi\hbar\Delta t}{m(x_b - x_a)^2} = \beta. \tag{B3}$$

The $(x_b - x_a)^2$ term in the denominator is simply $\delta^2$ and therefore,

$$\delta = \sqrt{\frac{2\pi\hbar\Delta t}{m\beta}}. \tag{B4}$$

From this expression, it is apparent that $\delta \sim \sqrt{\frac{\Delta t}{m}}$ for fixed β. Recall that Δt is found numerically from the approximation of the action integral and β can be assumed to be a constant independent of mass that determines when $K_{1step}$ has approximately constant wavelength oscillations. From numerical testing of a free particle with mass 1, we found that δ = 8 and Δt = 0.05 are sufficient for a numerically exact answer. This results in β = 0.005. Therefore, as mass increases, δ decreases such that the range of integration also decreases. Also, as Δt decreases, δ decreases and a smaller integration range may be used. It is in this manner that the PIQTr method is able to take advantage of a decreasing de Broglie wavelength for increasing mass. Table B1 shows the predicted values of δ from Eq. (B4) compared with the values found from numerical testing. While not exact, Eq. (B4) serves as a guide to estimate δ, generally to within a factor of 2.

| Mass | δ estimate from (17) | δ found numerically |
|------|----------------------|---------------------|
| 10   | 3                    | 3                   |
| 100  | 1.1                  | 2                   |
| 1000 | 0.79                 | 1.5                 |
| 10000| 0.35                 | 0.5                 |

**Table B1 Comparison of necessary integration range as predicted from Eq. (B4) with results obtained from numerical testing. Note that a mass of 1 was used to find β, which was in turn used to calculate δ, and therefore the two values must be identical. It was therefore excluded from the table.**

## Appendix C

One of the major challenges with the numerical calculation of heavy particle wave functions is the numerical and computational difficulty associated with increasing mass. It is therefore important that we understand how the computational requirements scale with increasing mass. We show in Table C1 the effect of mass on the computational requirements of the PIQTr method for a particle moving under a constant force. The values of Δt, $N_{xint}$, $N_{xwf}$, and δ that are required for accurate results with different masses are listed. A force of 2 a.u. was used for all masses, and all wave functions were propagated to the same final position ($x_f = 20$ a.u.) using the same initial state wave function.

| Mass | Maximum $\Delta t$ | Minimum $N_{xwf}$ | Minimum $N_{xint}$ | Minimum $\delta$ | Runtime per time step (s) |
|---|---|---|---|---|---|
| 1 | 0.05 | 10,000 | 2,000 | 8 | 0.55 |
| 10 | 0.07 | 15,000 | 10,000 | 3 | 0.67 |
| 100 | 0.1 | 50,000 | 10,000 | 2 | 1.74 |
| 1000 | 0.5 | 60,000 | 60,000 | 1.5 | 9.8 |
| 10000 | 1 | 100,000 | 80,000 | 0.5 | 15.5 |

**Table C1 Numerical requirements of the PIQTr method for different mass particles. Results are for a particle moving under a constant force of 2 a.u. All results were calculated on a single processor.**

From Table C1 it can be seen that the runtime per iteration changes by about a factor of 30 as the mass of the particle increases by 4 orders of magnitude. The primary parameters that influence the runtime are $N_{xwf}$ and $N_{xint}$. However, the minimum value for these parameters is coupled to the mass, $\Delta t$, and the range of integration $\delta$. A larger mass or smaller $\Delta t$ leads to greater oscillations in $K_{1step}$, which will require a greater density of $N_{xint}$ and $N_{xwf}$ points. However, a smaller $\delta$ for the same number of $N_{xint}$ and $N_{xwf}$ points leads to a greater point density. Therefore, as mass increases or $\Delta t$ decreases, we are able to keep $N_{xint}$ and $N_{xwf}$ at manageable values due to the decreasing $\delta$ required. We note that while it is advantageous for the runtime per iteration to remain relatively stable, a particle with larger mass will require a larger final time to reach the same final position. Therefore, the total computational time for the larger mass particles is increased over the smaller mass particles. However, the required minimum value for $\Delta t$ increases as mass increases, which helps keep runtimes down, even for large mass particles.

**Acknowledgements**

We gratefully acknowledge the support of the NSF under Grant Nos. PHY-1505217 and PHY-1838550.

# References

[1] H. He, M. E. J. Friese, N. R. Heckenberg, and H. Rubinsztein-Dunlop, *Phys. Rev. Lett.* **75**, 826 (1995).
[2] A. T. O'Neill, I. Mac Vicar, L. Allen, and M. J. Padgett, *Phys. Rev. Lett.* **88**, 053601 (2002).
[3] S. Fürhapter, A. Jesacher, S. Bernet, and M. Ritsch-Marte, *Opt. Express* **13**, 689 (2005).
[4] M. Baranek and Z. Bouchal, *J. Eur. Opt. Soc.* **8**, 13017 (2013).
[5] P. Galajda and P. Ormos, *Appl. Phys. Lett.* **78**, 249 (2001).
[6] P. Schattschneider, B. Schaffer, I. Ennen, and J. Verbeeck, *Phys. Rev.* B **85**, 134422 (2012).
[7] A. Asenjo-Garcia and F. J. Garcia de Abajo, *Phys. Rev. Lett.* **113**, 066102 (2014).
[8] M. Babiker, C. R. Bennett, D. L. Andrews, and L. C. Davila Romero, *Phys. Rev. Lett.* **89**, 143601 (2002).
[9] J. Verbeeck, H. Tian, and G. Van Tendeloo, *Adv. Mater.* **25**, 1114 (2013).
[10] J. Rusz and S. Bhowmick, *Phys. Rev. Lett.* **111**, 105504 (2013).
[11] S. Lloyd, M. Babiker, and J. Yuan, *Phys. Rev. Lett.* **108**, 074802 (2012).
[12] J. Verbeeck, H. Tian, and P. Schlattschneider, *Nature* **467**, 301 (2010).
[13] M. Uchida and A. Tonomura, *Nature* **464**, 737 (2010).
[14] B. J. McMorran, A. Agrawal, I. M. Anderson, A. A. Herzing, H. J. Lezec, J. J. McClelleand, and J. Unguris, *Science* **331**, 192 (2011).
[15] H. M. Scholz-Marggraf, S. Fritzsche, V. G. Serbo, A. Afanasev, and A. Surzhykov, *Phys. Rev.* A 90, 013425 (2014).
[16] M. V. Berry and N. L. Balazs, *Am. J. Phys.* **47**, 264 (1979).
[17] G. A. Siviloglou, J. Broky, A. Dogariu, and D. N. Christodoulides, *Phys. Rev. Lett*. 99, 213901 (2007).
[18] G. A. Siviloglou and D. N. Christodoulides, *Opt. Lett.* **32**, 979 (2007).
[19] J. Baumgartl, M. Mazilu, and K. Dholakia, *Nat Photon* **2**, 675–678 (2008).
[20] P. Rose, F. Diebel, M. Boguslawski, and C. Denz, *Applied Physics Letters* **102**, 101101 (2013).
[21] S. Jia, J. C. Vaughan, and X. Zhuang, *Nat Photon* **8**, 302–306 (2014).
[22] J. Broky, G. A. Siviloglou, A. Dogariu, and D. N. Christodoulides, *Optics Express* **16**, 12880 (2008).
[23] J. Bar-David, N. Voloch-Bloch, N. Mazurski, and U. Levy, *Nature Scientific Reports* **6**, 34272 (2016).
[24] Y. Zhang, H. Zhong, M. R. Belic and Y. Zhang, *Applied Science* **7**, 341 (2017).
[25] N. K. Efremidis and D. N. Christodoulide, *Optics Letters* **35**, 4045 (2010).

[26] D. Abdollahpour, S. Suntsov, D. G. Papazoglou, and S. Tzortzakis, *Phys. Rev. Lett.* **105**, 253901 (2010).
[27] P. Vaveliuk, A. Lencina, J. A. Rodrigo, and O. Martinez Mato, *Optics Letters* **39**, 2370 (2014).
[28] A. Patsyk, M. A. Bandres, R. Bekenstein, and M. Segev, *Phys. Rev.* X **8**, 011001 (2018).
[29] N. K. Efremidis and D. N. Christodoulides, *Opt. Lett.* **35**, 4045 (2010).
[30] N. K. Efremidis, V. Paltoglou, and W. von Klitzing, *Phys. Rev.* A **87**, 043637 (2013).
[31] S. Liu, M. Wang, P. Li, P. Zhang, and J. Zhao, *Opt. Lett.* **38**, 2416 (2013).
[32] Yiqi Zhang, Milivoj R. Belić, Huaibin Zheng, Haixia Chen, Changbiao Li, Yuanyuan Li, and Yanpeng Zhang, *Optics Express* **22**, 7160 (2014).
[33] Rui-Pin Chen, Chao-Fu Yin, Xiu-Xiang Chu, and Hui Wang, *Phys. Rev.* A **82**, 043832 (2010).
[34] N. Voloch-Bloch, Y. Lereach, Y. Lilach, et al., *Nature* **494**, 331 (2013).
[35] R. Shiloh, Y. Tsur, R. Remez, Y. Lereah, B. A. Malomed, V. Shvedov, C. Hnatovsky, W. Krolikowski, and A. Arie, *Phys. Rev. Lett.* **114**, 096102 (2015).
[36] B. Nelson, *Phys. Rev.* D **27**, 841 (1983).
[37] M. Winterstetter and W. Domcke, *Phys. Rev.* A **47**, 2838 (1993).
[38] M. Winterstetter and W. Domcke, *Phys. Rev.* A **48**, 4272 (1993).
[39] V. V. Smirnov, *Phys. Rev.* A **76**, 052706 (2007).
[40] R. Rosenfelder, *Phys. Rev.* A **79**, 012701 (2009).
[41] R. Rosenfelder, *J. Math. Phys.* **55**, 032106 (2014).
[42] E. B. Gravador, M. V. Carpio-Bernido, and C. C. Bernido, *Physics Letters* A **264**, 45 (1999).
[43] C. Gerry and V. Singh, *Phys. Rev.* D **21**, 2979 (1980).
[44] A. N. Vasil'ev and A. V. Kuz'menko, *Theor. Math. Phys.* **31**, 479 (1977).
[45] R. P. Feynman and A. R. Hibbs, Quantum Mechanics and Path Integrals, Emended Edition, Dover Publications (2005).
[46] Y. Fattal, A. Rudnick, and D. M. Marom, *Opt. Express* 19, 17298 (2011).
[47] Y. Zhang, M. Belić, Z. Wu, H. Zheng, K. Lu, Y. Li, and Y. Zhang, *Opt. Lett.* **38**, 4585 (2013).
[48] P. Panagiotopoulos, D. Abdollahpour, A. Lotti, A. Couairon, D. Faccio, D. G. Papazoglou, and S. Tzortzakis, *Phys. Rev.* A **86**, 013842 (2012).
[49] I. Kaminer, M. Segev, D. N. Christodoulides, *Phys. Rev. Lett.* **106**, 213903 (2011).
[50] Y. Hu, S. Huang, P. Zhang, C. Lou, J. Xu, and Z. Chen, *Opt. Lett.* **35**, 3952–3954 (2010).
[51] A. Rudnick and D. M. Marom, *Opt. Express* **19**, 25570 (2011).
[52] K. E. Strecker, G. B. Partridge, A. G. Truscott, and R. G. Hulet, *Nature* **417**, 150 (2002).
[53] F. Dalfovo, S. Giorgini, L. P. Pitaevskii, and S. Stringari, *Rev. Mod. Phys.* **71**, 463 (1999).